&biglatex
\documentclass[twocolumn]{rbef_editado}
\usepackage{xcolor}
\usepackage{float}
\usepackage{stmaryrd}
\usepackage{bbm}
\usepackage{subfig}
\usepackage{physics}
\usepackage{indentfirst}
\usepackage{subfig}
\usepackage{graphics, epstopdf, epsfig}
\usepackage{enumitem}
\usepackage{url}
\usepackage[hyphenbreaks]{breakurl}

\usepackage{hyphenat}
\newcommand{\daga}{^{\dagger}}
\newcommand{\operador}[1]{\mathbf{#1}}

\newcommand{\im}{\mathbbm{i}}
\newcommand{\Cmath}{\mathbbm{C}}

\DeclareMathAlphabet{\pazocal}{OMS}{zplm}{m}{n}
\newcommand{\Hcal}{\pazocal{H}}

\newcommand{\veryshortarrow}[1][3pt]{\mathrel{%
   \hbox{\rule[\dimexpr\fontdimen22\textfont2-.2pt\relax]{#1}{.4pt}}%
   \mkern-4mu\hbox{\usefont{U}{lasy}{m}{n}\symbol{41}}}}

\newcommand{\cnot}[2]{\mathbf{C}_\mathbf{x}^{(\text{#1}\veryshortarrow \text{#2})}}

\titulocabecalho{Algoritmos quânticos com IBMQ Experience}
\autorcabecalho{A.N. Oliveira, E.V.B. de Oliveira \textit{et al}}

\numeracao{xxxx}
\volume{xx}
\numero{xx}
\ano{2021xx}
\doi{http://dx.doi.org/xxxx}
\tipodeartigo{Artigos Gerais}

\author{Antonio N. Oliveira\thanks{Ambos autores contribuíram igualmente para este manuscrito}
}

\author{Estêvão V.B. de Oliveira$^\text{{\scriptsize $\ast$}}$\thanks{Correspondências devem ser enviadas para o endereço de e-mail: \href{mailto:evboliveira@df.ufscar.br}{evboliveira@df.ufscar.br}}~}

\author{Alan C. Santos
}

\author{Celso J. Villas-Bôas
}

\affil{Departamento de Física, Universidade Federal de São Carlos, 

Rodovia Washington Luís, km 235 - SP-310, 13565-905 São Carlos, SP, Brazil
}

\titulo{Algoritmos quânticos com IBMQ Experience: \\ Algoritmo de Deutsch-Jozsa}

\subtitulo{Quantum Algorithms in IBMQ Experience: Deutsch-Jozsa algorithm}

\begin{document}

\begin{primeirapagina}


\begin{abstract}
	

Processamento de informação quântica tem sido um dos pilares da nova era da informação. Nessa direção, o controle e processamento de informação quântica desempenha um papel fundamental, e computadores capazes de manipular tais informações tem se tornado realidade. Neste artigo nós apresentamos, de forma didática, elementos básicos da versão mais recente do computador quântico da IBM e suas ferramentas. Nós ainda apresentamos em detalhes o algoritmo de Deutsch-Jozsa usado para diferenciar funções constantes de funções balanceadas,  incluindo uma discussão de sua eficiência frente aos algoritmos clássicos para a mesma tarefa. A implementação experimental do algoritmo em um sistema de 4 qbits é apresentada. Nosso artigo abre caminho para uma série de investigações didáticas sobre o sistema da IBM, bem como os algoritmos quânticos mais conhecidos.
    \palavraschave{Computação quântica, algoritmos quânticos, empresa IBM, IBM-Q experience.}

	\end{abstract}
	
	\begin{otherlanguage}{english}
	
	\begin{abstract}
Quantum information processing has been one of the pillars of the new information age. In this sense, the control and processing of quantum information plays a fundamental role, and computers capable of manipulating such information have become a reality. In this article we didactically present basic elements of the latest version of IBM's quantum computer and its main tools. We also present in detail the Deutsch-Jozsa algorithm used to differentiate constant functions from balanced functions, also, including a discussion of its efficiency against classical algorithms for the same task. The experimental implementation of the algorithm in a 4-qbit system is presented. Our article paves the way for a series of didactic investigations into the IBM system as well as the best known quantum algorithms.
	\keywords{Quantum computation, quantum algorithms IBM company, IBM-Q Experience.}
	
	\end{abstract}
	\end{otherlanguage}

	\end{primeirapagina}
\saythanks

\section{Introdução}
\label{Sec:introducao}
O conceito de computação quântica se refere à utilização e manipulação de sistemas físicos regidos pela mecânica quântica para se realizar computação, de forma que a principal diferença entre um computador quântico e um computador clássico reside justamente na capacidade do primeiro de se favorecer das propriedades quânticas, como superposição, interferência e emaranhamento de estados, na busca por maior eficiência computacional. Assim, a origem da computação quântica nos remete à duas frentes, até então desassociadas, do conhecimento científico: a mecânica quântica e a ciência da computação. 

A física quântica datada do início do século XX, com os trabalhos de Max Planck~\cite{Planck1901}, se estabelece definitivamente como teoria moderna nos anos 1920, alicerçada nos trabalhos de Werner Heisenberg~\cite{Heisenberg1925} e, em paralelo, Erwin Schr\"odinger~\cite{Schrodinger1926} sobre a mecânica e cinemática de sistemas físicos como átomos e moléculas. No entanto, o avanço experimental para controle preciso de um único sistema quântico, individualmente, surge apenas em 1970 com a tecnologia de aprisionamento de átomos~\cite{atom_confined, Steane_1997, Lase_atoms}. Tal avanço foi fundamental para que hoje se fizesse possível alcançar a manipulação e controle fino de sistemas quânticos para a computação.

Paralelamente, em 1936, a ciência da computação dava seus passos iniciais com os trabalhos de Alan Turing, mostrando a existência de uma máquina universal que realiza operações tendo por base uma lógica de algoritmos. A partir dai, diversos desafios ao conceito de ``máquina de Turing'' (termo cunhado em sua homenagem) foram propostos, notadamente por R. Solovay e V. Strassen, mostrando a possibilidade do uso de algoritmos probabilísticos, superando o paradigma determinístico~\cite{Nielsen}.

A computação quântica, então, nasce de fato no início da década de 1980 quando Paul Benioff propõe um modelo de computador regido por um sistema quântico microscópico, e que satisfaz o conceito de uma máquina de Turing~\cite{Benioff1980}. Desde seu surgimento, uma das pautas centrais de debate a respeito do paradigma da computação quântica é a sua eficiência frente à computação clássica, e se um computador quântico poderia resolver problemas que não tinham uma solução eficiente na computação clássica.

Em 1985 David Deutsch respondeu positivamente a esta pergunta~\cite{DeutschDavid}, e nessa mesma linha, em 1994, Peter Shor mostrou que um computador quântico pode ser útil para a solução de grandes problemas, como a decomposição de um número inteiro em fatores primos, operação que se torna inviável em computadores clássicos quando o número a ser fatorado contém muitos dígitos~\cite{Shor_1997}. Ainda, um outro marco da área se deu em 1995 com a descoberta,  por Lov Grover, de que era possível, utilizando a computação quântica, conduzir uma pesquisa em uma base de dados não estruturada em uma velocidade maior que a desempenhada pelos algoritmos clássicos\cite{Grover}.

Ao lado de todo esse desenvolvimento, uma outra área da computação quântica, sugerida por Richard Feynman em 1982, também se desenvolveu a partir da ideia se usar os computadores quânticos para simular certos sistemas físicos que não poderiam ser simulados em computadores clássicos~\cite{Feynman1982}.

Neste artigo, o primeiro de uma série sobre algoritmos quânticos, nós apresentamos as ferramentas matemáticas que são a base para o entendimento da construção de algorítimos quânticos e implicações físicas e tecnológicas desses. Traduziremos a complexidade desses algoritmos para uma versão de simples entendimento, ampliando assim a acessibilidade ao público não especialista. Em particular, teremos como foco a descrição do Algoritmo de Deutsch-Jozsa~\cite{DeutschJozsa1992,deutsch-implementation} e sua implementação experimental, utilizando os computadores quânticos da IBM, disponíveis na plataforma IBMQ-Experi\-ence \cite{IBMQ} e de acesso ao público, para programação na nuvem. O IBMQ-Experience tem sido usado para implementar tarefas quânticas, como por exemplo teletransporte~\cite{AlanBloch,Rabelo:18}, geração de estados emaranhados~\cite{Jesus:21}, e simular sistemas quânticos~\cite{Alves:20}.

\subsection{Ferramental matemático da Mecânica Quântica}
\label{sec:introdução_a_mec_quantica}

Esta seção será destinada a uma breve apresentação dos fundamentos e conceitos básicos que serão usados ao longo do artigo, sendo, assim, indispensáveis para o entendimento das seções que se seguem.

O primeiro conceito a ser introduzido é a ideia da representação do estado de um sistema físico através de um vetor $\ket*{\psi}$, em notação \textit{Bra-ket}\footnote{A notação \textit{Bra-ket} foi proposta por Paul Dirac e por isso também pode ser referida como notação de Dirac. É muito utilizada no campo da mecânica quântica para representar vetores, produtos internos e produtos externos. O \textit{Ket} é utilizado para representar um vetor
    $$\ket*{\psi}= \mqty[\psi_{1}&\psi_{2}&...&\psi_{n}]^T.$$
Já o \textit{Bra} representa um vetor no espaço dual de $\ket*{\Psi}$, sendo um vetor linha cujos elementos são os complexos conjugados de $\ket*{\Psi}$
    $$\ket*{\psi}= \mqty[\psi_{1}^{*}&\psi_{2}^{*}&...&\psi_{n}^{*}] \, .$$}, que reside em um espaço vetorial complexo (corpo dado pelos números complexos) composto de vetores ortogonais e normalizados (norma igual à $1$). A esse espaço, dá-se o nome de \textit{espaço de Hilbert do sistema} e denota-se como $\Hcal$. Tal espaço vetorial é munido de um produto interno e relação de completeza
~\cite{arfken}. O produto interno entre dois elementos $\ket*{\psi},\ket*{\phi}\in\Hcal$ escreve-se $\braket{\psi}{\phi}\!=\!(\ket*{\psi},\ket*{\phi})$, este sendo um número definido no corpo de $\Hcal$, ou seja $\braket{\psi}{\phi}$ é um número complexo ($\Cmath$), em geral.

No contexto de computação quântica, os sistemas físicos de interesse são aqueles que podem ser representados por vetores de estado pertencentes a um espaço de Hilbert bidimensional, caracterizando um sistema quântico de dois níveis, o \textit{qbit}. A base ortonormal utilizada na representação dos vetores de estado é a base canônica $\{\ket*{0}, \ket*{1}\}$, com representação matricial
\begin{align}
    \ket*{0} = \mqty[1\\0] 
    \, ,
    \hspace{1cm}
    \ket*{1} = \mqty[0\\1]
    \, .
\end{align}
Doravante, a base acima será referida como \textit{base computacional}, sendo $\ket*{0}$ e $\ket*{1}$ os \textit{estados da base computacional}. Portanto, qualquer vetor de estado $\ket*{\psi}$ pode ser, então, escrito como
\begin{align}
\ket*{\psi}=a\ket*{0}+b\ket*{1}
\, , \label{eq:estado_superposicao}
\end{align}
um estado de superposição entre os vetores da base, em que os coeficientes complexos $a$ e $b$ são denominados amplitudes de probabilidade. Da normalização do vetor de estado, segue que, a partir dessa representação, os coeficientes $a$ e $b$ obedecem à condição $|a|^2+|b|^2=1$.

Como objetos que atuam sobre tais vetores, definimos operadores sobre o mesmo espaço vetorial, $\operador{O}\in\Hcal$,  tal que
\begin{align}
    \operador{O}\ket*{\psi}\!=\!\ket*{\psi'} \, ,
\end{align}
com $\ket*{\psi'}\!\in\!\Hcal$. Além disso, operadores em $\Hcal$ geram transformações lineares em $\Hcal$, ou seja, para quaisquer $\ket*{\phi}$,$\ket*{\psi}$ $\in\Hcal$ e $a,b\in\Cmath$ vale
\begin{align}
    \operador{O} \left(a\ket*{\phi}+b\ket*{\psi}\right) =& a ~\operador{O}  \ket*{\phi}+b~\operador{O} \ket*{\psi}
    \nonumber \\
    =& a  \ket*{\phi'}+b\ket*{\psi'}
    \, .
\end{align}

Dado o conjunto infinito de operadores, destacamos as matrizes de Pauli $\operador{X}$, $\operador{Y}$ e $\operador{Z}$; matrizes $2$x$2$ que, junto à matriz identidade $\operador{I}$, constituem um conjunto que descreve operações em sistemas quânticos de dois níveis, e por isso são de grande importância no campo da computação quântica. Essas matrizes, e o modo com que elas transformam os estados da base computacional são
\begin{subequations}
\begin{align}
    \operador{X} = \mqty[0&1\\1&0]
    &\implies
    \left\{
    \begin{aligned}
    \operador{X} \ket*{0}&=\ket*{1}\\
    \operador{X}  \ket*{1}&=\ket*{0}
    \end{aligned}
    \right.
    \, , \label{definicao_X} \\
    \operador{Y} = \mqty[0&-\im\\\im&0]
    &\implies
    \left\{
    \begin{aligned}
    \operador{Y} \ket*{0}&= \im\ket*{1}\\
    \operador{Y}  \ket*{1}&= -\im\ket*{0}
    \end{aligned}
    \right.
    \, , \label{definicao_Y} \\
    \operador{Z} = \mqty[1&0\\0&-1]
    &\implies
    \left\{
    \begin{aligned}
    \operador{Z}  \ket*{0}&= \ket*{0}\\
    \operador{Z} \ket*{1}&= -\ket*{1} 
    \end{aligned}
    \right.
    \, ,\label{definicao_Z}
\end{align}
\end{subequations}
em que $\im\!=\!\sqrt{-1}$ é a unidade imaginária.

Ainda, a evolução temporal de um sistema quântico fechado é dada através do operador de evolução temporal $\operador{U}(t)$, ou seja, um operador quântico que leva um vetor de estado $\ket*{\psi(0)}$ a um vetor de estado $\ket*{\psi(t)}$. isto é, 
\begin{align}
\label{eq:evol_temporal}
    \ket*{\psi(t)} = \operador{U}(t) \ket*{\psi(0)}
    \, ,
\end{align}
sendo $\operador{U}(t)$ determinado pela equação de Schr\"odinger \cite{sakurai2013mecanica}.
Uma importante propriedade do operador de evolução temporal $U(t)$ é a sua unitariedade, ou seja, 
\begin{align}
\label{eq:ope unitaria}
\operador{U}\daga \cdot \operador{U} = \operador{I}
\, ,
\end{align}
em que $\operador{U}\daga$ é o hermitiano conjugado de $\operador{U}$. A implicação direta é
\begin{align}
\label{eq:ope unitaria norma}
    \braket{\psi(t)}{\psi(t)} 
    &= \bra{\psi(0)}\,\operador{U}\daga(t) \cdot \operador{U}(t)\,\ket*{\psi(0)} 
    \nonumber \\
    &= \bra{\psi(0)}\, \operador{I} \,\ket*{\psi(0)} 
    \nonumber \\ 
    &=\braket{\psi(0)}{\psi(0)} = 1
    \, ,
\end{align}
preservando a condição de normalização. 

\subsection{Sistemas quânticos compostos}

Em computação, sempre temos que lidar com muitos dados independentes. Então, sistemas quânticos devem ser capazes de funcionar sob essas condições. Nesse sentido, é sempre possível gerar estados compostos, ou seja, estados quânticos que representam o estado \textit{conjunto} de dois ou mais qbits. Matematicamente isso é feito via produto tensorial e denotado como
\begin{align}
    \ket*{\Psi} = \ket*{\psi_1} \otimes \ket*{\psi_2} \otimes  \dots  \otimes \ket*{\psi_n} \label{EqTensorPsi}
    \, ,
\end{align}
sendo $\ket*{\psi_i}$ o vetor de estado relativo ao $i$-ésimo qbit com seu espaço de Hilbert $\Hcal_{i}$ independente. Junto com o estado, o espaço de Hilbert total é também aumentado, sendo $\ket*{\Psi}\!\in\!\Hcal_{\mathrm{total}}\!=\!\Hcal_{1}\otimes \Hcal_2 \otimes  \dots  \otimes \Hcal_n$. Analogamente, uma composição de um conjunto de operações é escrita da seguinte forma
\begin{align}
    \operador{O}_T = \operador{O}_1 \otimes \operador{O}_2 \otimes  \dots  \otimes \operador{O}_n = \bigotimes_{i=1}^{n} \operador{O}_i
    \, ,
\end{align}
Onde $\operador{O}_i$ são operadores que atuam no espaço de Hilbert do $i$-ésimo qbit. O símbolo de produtória tensorial ``$\bigotimes$'' é usado para representar o protudo de uma forma compacta. A atuação do operador $\operador{O}_T$ no estado $\ket*{\Psi}$ definido na Eq.~\eqref{EqTensorPsi} é dada por
\begin{align}
    \ket*{\Psi'}&=\operador{O}_T\ket*{\Psi} \nonumber \\
    &= \left(\operador{O}_1 \otimes \operador{O}_2 \otimes  \dots  \otimes \operador{O}_n\right) \left(\ket*{\psi_1} \otimes \ket*{\psi_2} \otimes  \dots  \otimes \ket*{\psi_n}\right)
    \nonumber \\
    &=\left(\operador{O}_1\ket*{\psi_1}\right) \otimes \left(\operador{O}_2\ket*{\psi_2}\right) \otimes  \dots  \otimes \left(\operador{O}_n\ket*{\psi_n}\right)
    \nonumber \\
    &= \ket*{\psi_1'} \otimes \ket*{\psi_2'} \otimes  \dots  \otimes \ket*{\psi_n'}
    \, .
\end{align}
Para alívio de notação, também é possível omitir-se o símbolo ``$\otimes$'' na representação da composição de espaços vetoriais para os vetores de estado $\ket*{\psi_i}$, escrevendo-se $\ket*{\psi_1}\ket*{\psi_2} \dots \ket*{\psi_n}$ ou simplesmente $\ket*{\psi_1,\psi_2, \dots ,\psi_n}$.

\subsection{Elementos de computação quântica}
\label{sec:intro_comp_quantica}

Além das matrizes de Pauli, existe uma series de operações importantes no contexto da computação quântica, sendo elas atuantes em $1$ ou mais qbits. Uma relevante porta lógica quântica de $1$ qbit é a porta Hadamard, cuja representação matricial e atuação são
\begin{align}
	\label{eq:definicao_Hadamard}
	\operador{H}=\dfrac{1}{\sqrt{2}} \mqty[1&1\\1&-1]
	\implies
	\operador{H}\ket*{x}=\frac{\ket*{0}+(-1)^x\ket*{1}}{\sqrt{2}}
	\, ,
\end{align}
ou simplesmente,
\begin{align}
	\operador{H} \ket*{x}
	=
	\left\{
	\begin{matrix}
		\frac{\ket*{0} + \ket*{1}}{\sqrt{2}}=\ket*{+} & \text{ , se } x=0 \\
		&\\
		\frac{\ket*{0} - \ket*{1}}{\sqrt{2}}=\ket*{-} & \text{ , se } x=1
	\end{matrix}
	\right.
	\, .
\end{align}
que resulta na criação de um estado de superposição entre os vetores de estado da base computacional com igual amplitude de probabilidade.

Definindo as operações de $2$ qbits temos a porta lógica CNOT, ou ``não controlado'' $\cnot{c}{t}$, em que $c$ representa o qbit de controle e $t$ o qbit alvo. Sua atuação muda o estado do qbit alvo dependendo do valor do qbit de controle	
\begin{align}
	\cnot{1}{2}=
	\mqty[1&0&0&0\\0&1&0&0\\0&0&0&1\\0&0&1&0]
	\implies
	\left\{
	\begin{matrix}
		\ket*{00}\rightarrow\ket*{00}\\
		\ket*{01}\rightarrow\ket*{01}\\
		\ket*{10}\rightarrow\ket*{11}\\
		\ket*{11}\rightarrow\ket*{10}
	\end{matrix}
	\right.
	\,.\label{definicao cnot}
\end{align} 

Todas as portas quânticas podem ser resumidas a um conjunto limitado de portas de 1 e 2 qbits. Esse conjunto envolve as chamadas portas Clifford (porta Hadamard, portas de Pauli, a CNOT e a porta de fase $S$) somadas às portas de rotação ($R_y$ e $R_z$) \cite{Nielsen}, outras portas serão detalhadas nos próximos artigos dessa série.

Na computação clássica, a sequência de portas lógicas atuando em um sistema pode ser representada na forma de um circuito e a unidade lógica é definida como o \textit{bit}, que assume os valores $0$ ou $1$. De forma análoga, na computação quântica um circuito quântico é uma maneira de caracterizar o sequenciamento das operações quânticas que atuam na unidade lógica, o \textit{qbit}, com a ressalva de que, nesta, as propriedades quânticas de superposição e emaranhamento surgem como um gran\-de diferencial com relação à computação clássica, conforme será detalhado nas seções seguintes.

\begin{figure}
	\centering \includegraphics[width = 0.7 \columnwidth]{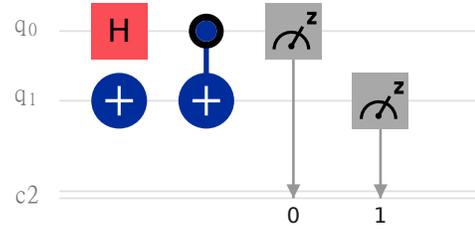}
	\caption{Esquema representativo de um circuito quântico, em que as linhas simples representam os registros quânticos, e as linhas duplas representam os registros clássicos. Cada bloco representa uma operação quântica, e ao final do circuito realiza-se as medidas dos qbits $q_0$ e $q_1$, armazenando-se os resultados no registro clássico. Na seção \ref{sec:ibmq_experience} será detalhada a plataforma utilizada para a construção deste circuito, a IBMQ experience, bem como qual operação quântica é representada por cada bloco.}
	\label{fig:esquema_circuito_quântico}
	\vspace{-0.4cm}
	\begin{flushleft}
		{\small Fonte: IBM Q~\cite{IBMQ}.}
	\end{flushleft}
\end{figure}

A figura \ref{fig:esquema_circuito_quântico} mostra o esquema de um circuito quântico, em que cada linha representa um qbit, e a leitura é feita da esquerda para a direita. Na figura mencionada, $q$ é o registro quântico contendo dois qbits, $q_0$ e $q_1$, em que se implementam as operações, sendo que, quando os operadores estão alinhados verticalmente temos a atuação simultânea, no tempo, desses operadores nos seus respectivos qbits. Já o elemento $c_2$ representa um registro clássico de $2$ bits, que irá armazenar a informação resultante de uma medida dos qbits, podendo assumir os valores $00$, $01$, $10$ ou $11$. Ao final do circuito é realizada, então, a operação de medição, definida como sendo a medição dos qbits na base computacional, ou seja, a base Z.

Além disso, uma importante propriedade dos circuitos quânticos é sua reversibilidade, isto é, devido à condição de unitariedade dos operadores quânticos, as operações em um circuito quântico podem ser revertidas simplesmente aplicando o mesmo operador seguidamen\-te, como no exemplo que segue para um estado $\ket*{\psi}=\ket*{0}$ e a operação Hadamard atuando sobre esse estado
\begin{align}
	 \ket*{0} &\xrightarrow{\text{Hadamard}} \frac{\ket*{0} + \ket*{1}}{\sqrt{2}} = \ket*{+}
	\nonumber \\
	\ket*{+}
	 &\xrightarrow{\text{Hadamard}} \frac{1}{\sqrt{2}} \left(\frac{\ket*{0} + \ket*{1}}{\sqrt{2}} + \frac{\ket*{0} - \ket*{1}}{\sqrt{2}}
	 \right)=\ket*{0} .
\end{align}

Por convenção um circuito é iniciado com todos os qbits no estado $\ket*{0}$ e a partir dai as portas lógicas quânticas são implementadas. Tais portas nada mais são do que operadores quânticos que atuam em um sistema composto por um ou mais qbits. 

\section{Plataforma IBMQ Experience}
\label{sec:ibmq_experience}
A computação quântica tem se desenvolvido nas últimas décadas e, cada vez mais, grandes empresas de tecnologia como Google, D-Wave, IonQ e IBM entram nesse setor em busca da comprovação da superioridade, de maneira geral, dos computadores quânticos frente aos clássicos, a chamada ``supremacia quântica''~\cite{IBM2020,google2021,ionq2020}. Neste trabalho nos restringimos ao uso da IBMQ Experience, uma plataforma para computação quântica disponibilizada pela empresa IBM~\cite{IBMQ}. Gratuita e com uma interface para computação em nuvem, nela temos acesso a diversos computadores quânticos que variam de $5$ a 15 qbits. Além disso é possível acessar os simuladores (computadores clássicos que simulam um computador quântico ideal) com capacidade de 32 a 5000 qbits. A plataforma permite a criação de circuitos tanto em uma interface gráfica, em um modelo de ``clique e arraste'' das portas lógicas, como a partir de linguagens de programação para computação quântica, notadamente a linguagem QASM~\cite{qasm} e a biblioteca Qiskit~\cite{Qiskit} para programação em linguagem Python.

\begin{figure}
\centering
\subfloat[Identidade]{\qquad \includegraphics[scale=1]{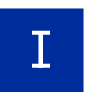}\qquad}
\qquad
\subfloat[Pauli X]{\quad\includegraphics[scale=.5]{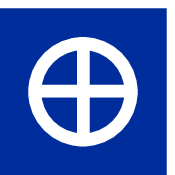}\quad}
\qquad
\subfloat[Hadamard]{\qquad \includegraphics[scale=1]{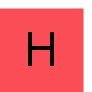}\qquad}
\\
\subfloat[CNOT]{\includegraphics[scale=.7]{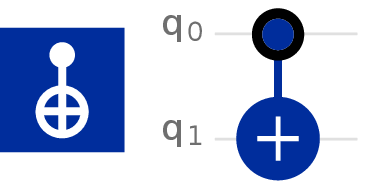}}
\qquad
\subfloat[Medidor clássico]{\qquad \includegraphics[scale=.6]{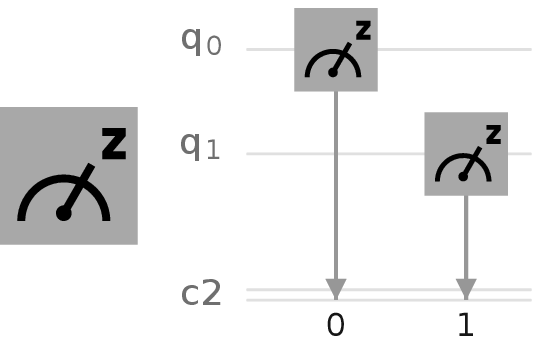} \qquad}
\caption{Ícones e representações referentes às portas lógicas quânticas de um e dois qbits utilizadas e à operação de medição na base computacional Z, disponíveis na IBM Quantum Experience.  A operação de medição converte o valor obtido para os estados quânticos dos qbits em bits, e armazena no registro clássico, em que o primeiro qbit (referente ao registro $q_0$) é o primeiro dígito (menor base de 2) do número binário no registro clássico, e assim por diante, de modo que o resultado da medição do estado $\ket*{q_0,q_1}=\ket*{1,0}$, por exemplo, seria o número binário $01$.}
\label{fig:ICONES_IBM}
\vspace{-0.4cm}
\begin{flushleft}
	{\small Fonte: IBM Q~\cite{IBMQ}  e editada pelos autores.}
\end{flushleft}
\end{figure}

Por meio da utilização da interface gráfica podemos clicar nos ícones que representam os operadores quânticos e posicioná-los no circuito para que atuem sobre um qbit e, desse modo, apliquem sobre eles suas operações matemáticas, como ilustra a figura~\ref{fig:esquema_circuito_quântico}. Na figura \ref{fig:ICONES_IBM} podemos ver alguns dos ícones disponíveis referentes às operações já mencionadas anteriormente. Após a construção do algoritmo, este pode ser submetido para processamento em um computador quântico. A possibilidade de escolha, para esse experimento na plataforma,  do número de "shots", contagem dos resultados de experimentos identicamente preparados, objetiva aumentar a acurácia estatística dos resultados. Ao final da execução, os resultados são dispostos na forma de um histograma que reflete as probabilidades das possíveis respostas. 

O dispositivo utilizado para as implementações descritas nesse artigo foi o IBMQ Manila, um computador quântico com 5 qbits, cuja topologia, que destaca as conexões entre esses qbits, é ilustrada pela figura 2. O número de ``shots'' escolhido para todas as execuções foi 8000. 

\begin{figure}
	\centering
	\includegraphics[width =  \columnwidth]{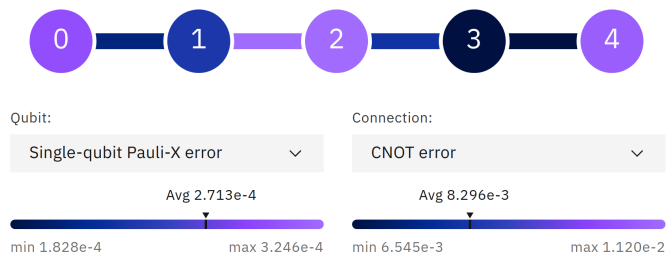}
	\caption{Topologia do computador quântico IBMQ Manila, em que cada círculo enumerado simboliza 1 qbit, e as ligações entre eles simbolizam as conectividades entre os qbits. A escala de cores representa os erros associados às operações quânticas realizadas neste computador, em que, para os qbits, essa escala está associada à taxa de erros da operação de rotação $\operador{X}$, e, para as conexões, está associada à taxa de erros da operação CNOT entre os qbits. Os valores mostrados estão de acordo com a calibração do computador no momento em que os experimentos foram realizados.}
	\label{fig:topologia IBM manila}
	\vspace{-0.4cm}
	\begin{flushleft}
	{\small Fonte: IBM Q~\cite{IBMQ}.}
	\end{flushleft}
\end{figure}

Por fim, cabe ainda a menção de que os computadores quânticos, sendo sistemas físicos não ideias, apresentam erros associados a processos de dissipação e decoerência, inerentes a um sistema quântico não perfeitamente isolado\footnote{Processos de dissipação e decoerência são resultados da interação de um sistema quântico com o ambiente externo, e têm como principal consequência a perda de algumas propriedades quânticas do sistema, como, por exemplo, a unitariedade das operações descritas de acordo com as Eqs. \eqref{eq:ope unitaria} e \eqref{eq:ope unitaria norma}, e a superposição coerente de estados descrita na Eq. \eqref{eq:estado_superposicao}. Uma discussão mais completa sobre esse aspecto dos computadores quânticos e da plataforma IBMQ é encontrada em~\cite{AlanBloch,TCCEstevao}}. Ainda, a plataforma fornece uma documentação com um rico manual de instruções para os usuários, disponível em~\cite{IBMQ_documentation}.

\section{O algoritmo de Deutsch}
\label{sec:alg_deutsch}

O algoritmo de Deutsch se propõe a utilizar da propriedade quântica da superposição coerente de estados, algo inexistente em dispositivos clássicos, para então determinar, a partir de uma única medida, se uma dada função $f(x)$ é constante ou balanceada.

Seja $f(x)$ uma função binária tal que $f: \{0,1\} \rightarrow \{0,1\}$. Se $f(x)$ é balanceada temos as opções $f(0)=0 \, , \, \,f(1)=1$ ou $f(0)=1 \, , \, \,f(1)=0$. Caso ela seja constante temos $f(0)=f(1)=0$ ou $f(0)=f(1)=1$. Para ambos os casos, em um computador clássico, seria necessária a avaliação da função $f(x)$ para todos os dois valores $x$ de entrada para se determinar com certeza se a função é constante ou balanceada. Já num computador quântico veremos que, devido às propriedades de superposição de estados quânticos, é possível que a função $f(x)$ seja avaliada para todos os dois valores de $x$ em paralelo e simultaneamente, e ainda, a partir da propriedade interferência quântica, é possível que seja determinado, com certeza, por meio de uma única medida, se $f(x)$ é constante ou balanceada.

\begin{figure}
	\centering
	\includegraphics[scale=0.5]{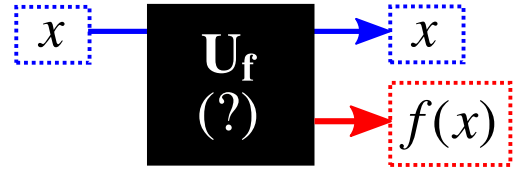}
	\caption{Esquema ilustrativo da atuação da operação $\operador{U_f}$, isto é, uma ``caixa preta'' que de alguma forma implementa a função $f$ em seu mecanismo interno. Quando é fornecido uma certa entrada $x$, a resposta obtida na saída é $f(x)$, ou seja, a função $f$ avaliada no valor $x$.}
	\label{fig:esquema_mecanismo_caixa_preta_deutsch}
	\vspace{-0.4cm}
	\begin{flushleft}
		{\small Fonte: Elaborado pelos autores.}
	\end{flushleft}
\end{figure}

Para a construção do algoritmo serão necessários dois registros quânticos de $1$ qbit, o primeiro para representar as entradas $x$ e o segundo para representar os valores possíveis da função $f(x)$, da forma $\ket*{x}\ket*{f(x)}$. Ainda, definimos a operação unitária $\operador{U_f}$ cuja atuação em um estado é dada por
\begin{align}
\label{eq:operacao_Uf}
    \operador{U_f} \ket*{x}\ket*{y} = \ket*{x} \ket*{y \oplus f(x)}
    \, ,
\end{align}
em que $x,y \in \{0,1\}$ e o símbolo ``$\oplus$'' representar a soma em módulo 2 \cite{IMPA}. No contexto do algoritmo de Deutsch, a atuação do operador $\operador{U_f}$ no estado $\ket*{x}$ representa a avaliação do valor da função $f$ para a entrada $x$, e armazenamento deste resultado no segundo registro quântico. Entretanto, diferente das operações quânticas já definidas nas seções \ref{sec:introdução_a_mec_quantica} e \ref{sec:intro_comp_quantica}, não nos preocuparemos, por enquanto, com a representação matricial do operador $\operador{U_f}$, nem com os detalhes do mecanismo interno por trás da implementação deste. Iremos tratá-lo apenas como uma ``caixa preta'' ou \textit{blackbox}, em inglês, cujo mecanismo interno, não importa qual, implementa a operação da Eq. \eqref{eq:operacao_Uf}, esquematizada na figura \ref{fig:esquema_mecanismo_caixa_preta_deutsch}. A única coisa que assumiremos será que a operação $\operador{U_f}$ nos é dada, e munidos dela, implementaremos o algoritmo de Deutsch para avaliar se a função $f(x)$ é constante ou balanceada. Nas seções que seguirão, discutiremos a respeito desta suposição, fornecendo mais detalhes acerca desta ``caixa preta''. 

Prosseguindo, será necessário um registro clássico de um bit para, ao final do algoritmo, armazenar o resultado de uma medição do registro de $1$ qbit que representa a função $f(x)$. O vetor de estado inicial é então iniciado com o primeiro qbit no estado $\ket*{0}$ e o segundo no estado $\ket*{1}$
\begin{align}
    \ket*{\psi^{(0)}} = \ket*{0}\ket*{1}
    \, .
\end{align}

Aplicando então a operação Hadamard em ambos os registros, colocamos ambos os qbits em um estado de superposição de todos os estados da base computacional ($\ket*{0}$ e $\ket*{1}$), como mostrado na Eq. \eqref{eq:definicao_Hadamard}, 
\begin{align}
    \ket*{\psi^{(1)}} 
    &= (\operador{H} \otimes \operador{H}) \ket*{\psi^{(0)}} = \left( \operador{H} \ket*{0}\right) \otimes \left(\operador{H} \ket*{1} \right)
    \nonumber \\
    &= \ket*{+}\ket*{-}
    \, .
\end{align}
Assim, aplicando a operação $\operador{U_f}$ em seguida temos o fenômeno do paralelismo quântico em ação, já que a função $f(x)$ será calculada pra $x=0$ e $x=1$, simultaneamente. Assim 
\begin{align}
\label{eq:deutsch_aplicacao_Uf}
    \ket*{\psi^{(2)}} 
    = \operador{U_f} \ket*{\psi^{(1)}} = \operador{U_f} \ket*{+} \ket*{-}
    \, .
\end{align}
Mas 
\begin{align}
    \operador{U_f} \ket*{x} \frac{\ket*{0}-\ket*{1}}{\sqrt{2}}
    &=\ket*{x} \frac{\ket*{0\oplus f(x)}-\ket*{1 \oplus f(x)}}{\sqrt{2}}
    \nonumber \\
    &=\left\{
    \begin{matrix}
    \ket*{x} \frac{\ket*{0}-\ket*{1}}{\sqrt{2}} & \text{ se } f(x)=0 \, , \\[0.3cm]
    \ket*{x} \frac{\ket*{1}-\ket*{0}}{\sqrt{2}} & \text{ se } f(x)=1 \, ,
    \end{matrix}
    \right.
\end{align}
implicando que
\begin{align}
    \operador{U_f} \ket*{x} \ket*{-}
    = (-1)^{f(x)} \ket*{x} \ket*{-}
    \, ,
\end{align}
Logo, a Eq. \eqref{eq:deutsch_aplicacao_Uf} fica
\begin{align}
    \ket*{\psi^{(2)}} 
    = \frac{(-1)^{f(0)}\ket*{0}+(-1)^{f(1)}\ket*{1}}{\sqrt{2}} \otimes\ket*{-}
    \, .
\end{align}
Por fim, aplicamos a operação Hadamard no primeiro registro
\begin{align}
\label{eq:deutsch_aplicacao_Uf_hadamard}
    \ket*{\psi^{(3)}} &= (\operador{H}\otimes \operador{I}) \ket*{\psi^{(2)}}
    \nonumber \\
    &=\left[(-1)^{f(0)} \frac{\ket*{0}+\ket*{1}}{2} + (-1)^{f(1)} \frac{\ket*{0}-\ket*{1}}{2}\right]\otimes\ket*{-}
    \, .
\end{align}
É nesta etapa que a interferência entre os estados quânticos atua no sistema, fazendo com que os estados que não representam a resposta esperada se interfiram destrutivamente, e que o estado cuja medida representa a resposta para o problema da função $f(x)$ tenha sua amplitude de probabilidade aumentada. Para que essa inferência se torne evidente precisamos analisar cuidadosamente os diferentes casos. Quando $f(x)$ for uma função \textit{constante} teremos $f(1)=f(0)= \alpha$, com $\alpha$ uma constante qualquer. Assim a Eq. \eqref{eq:deutsch_aplicacao_Uf_hadamard} fica
\begin{align}
\label{eq:deutsch_aplicacao_Uf_hadamard_constante}
    \ket*{\psi^{(3)}} &= (-1)^{\alpha} \left[ \frac{\ket*{0}+\ket*{1}}{2} + \frac{\ket*{0}-\ket*{1}}{2}\right]\otimes\ket*{-}
    \nonumber \\
     &= (-1)^{\alpha} \ket*{0}\otimes\ket*{-}
    \, .
\end{align}

Ainda, se $f(x)$ for uma função \textit{balanceada}, teremos $f(0)\neq f(1)$. Aqui, é possível verificar que, para esse caso,
$$f(1)=1\oplus f(0) \implies (-1)^{f(1)} = -(-1)^{f(0)} \, ,$$
seja $f(0)=0$ e $f(1)=1$, ou $f(0)=1$ e $f(1)=0$. Assim a Eq. \eqref{eq:deutsch_aplicacao_Uf_hadamard} fica
\begin{align}
\label{eq:deutsch_aplicacao_Uf_hadamard_balanceada}
    \ket*{\psi^{(3)}} &= (-1)^{f(0)} \left[ \frac{\ket*{0}+\ket*{1}}{2} - \frac{\ket*{0}-\ket*{1}}{2}\right]\otimes\ket*{-}
    \nonumber \\
    &= (-1)^{f(0)} \ket*{1}\otimes \ket*{-}
    \, .
\end{align}
Logo, das Eqs. \eqref{eq:deutsch_aplicacao_Uf_hadamard_constante} e \eqref{eq:deutsch_aplicacao_Uf_hadamard_balanceada} vemos que em uma única medição no primeiro registro conseguimos determinar, com certeza, a característica da função, com os seguintes resultados:
\begin{align}
    \left\{
    \begin{aligned}
        0 \, , & \text{ se $f(x)$ for constante,}\\
        1 \, , & \text{ se $f(x)$ for balanceada.}
    \end{aligned}
    \right.
\end{align}
De fato, ainda é possível reunir os resultados de ambas Eqs. \eqref{eq:deutsch_aplicacao_Uf_hadamard_constante} e \eqref{eq:deutsch_aplicacao_Uf_hadamard_balanceada} em uma única, nos atentando para o fato de que 
\begin{align}
f(0) \oplus f(1) =
    \left\{
    \begin{aligned}
        0 \, , & \text{ se } f(0)=f(1) \, , \\
        1 \, , & \text{ se } f(0)\neq f(1), 
    \end{aligned}
    \right.
\end{align}
e assim, a Eq. \eqref{eq:deutsch_aplicacao_Uf_hadamard} se resume ao resultado final
\begin{align}
    \ket*{\psi^{(3)}}
    &= (-1)^{f(0)} \ket*{f(0) \oplus f(1)} \otimes\ket*{-}
    \, ,
\end{align}
atestando que em uma única avaliação da função $f(x)$ obtemos sua informação global.

\section{Algoritmos de consulta e o mecanismo de ``caixa preta''}
\label{sec: caixa preta}
 Na seção anterior vimos que, utilizando o Algoritmo de Deutsch, é possível determinar se uma função binária de duas entradas $f(x): \{0,1\} \rightarrow \{0,1\}$ é constante ou balanceada, a partir da operação $\operador{U_f}$, em apenas uma avaliação dessa função, o que, num circuito quântico, é representado pelo ato da medida. Entretanto, nada foi mencionado à respeito da estrutura, representação matricial, ou mecanismo por trás desta operação, assumindo, apenas, que ela consistia em uma ``caixa preta'' que atua conforme a figura \ref{fig:esquema_mecanismo_caixa_preta_deutsch}.
 
 Nesse contexto, tal caixa preta pode ser entendida co\-mo um agente que, de algum modo, tem um conhecimento acerca da função $f(x)$, sendo, então, capaz de imple\-mentá-la a partir de algum mecanismo o qual ain\-da não conhecemos. Essa definição caracteriza um operador Oráculo, presente na categoria dos chamados  Algoritmos Quânticos de Consulta, ou \textit{Quantum Query Algorithms}, em inglês, tal como os já mencionados algoritmos de Shor e Grover. Nessa categoria de algoritmos o mecanismo de obtenção da resposta do problema inclui consultas a esse Oráculo. No algoritmo de Grover para busca em uma lista não estruturada por exemplo, este operador, a partir de uma certa entrada fornecida, nos devolve como saída, ``sim'', se a entrada é o elemento buscado, ou ``não'', caso contrário\footnote{Uma interessante analogia que ilustra bem este caso é o mecanismo ``chave-fechadura'', em que apesar de a fechadura (Oráculo), devido à sua estrutura interna, ``saber'' qual das chaves num molho (lista) a destranca, ela não nos aponta de prontidão qual é a correta, exigindo que façamos uma consulta para cada uma das chaves, individualmente -- no caso não quântico pelo menos. A fechadura apenas nos retorna ``sim'' (destrancando) ou ``não'' (emperrando) a respeito da chave com a qual fazemos a consulta.}. Esse comportamen\-to binário, é facilmente traduzido no contexto da computação, tanto clássica quanto quântica, e, para o operador Oráculo do algoritmo de Deutsch, a ``resposta'' é $0$ ou $1$, representando os possíveis valores da função binária dependendo da entrada fornecida na consulta. De certa forma, tal definição introduz uma série de dúvidas no que diz respeito à ``transparência'' do mecanismo de funcionamento do algoritmo de Deutsch, e de como imple\-mentá-lo na prática. 
 
Para exemplificar o ponto acima, suponha que tenhamos em mãos uma certa função, cuja característica (constante ou balanceada) pretendemos determinar a partir do algoritmo de Deutsch. Além da representação das variáveis $x$ e da função $f$ em vetores de estado, e da implementação das operações Hadamard, já detalhadas na seção \ref{sec:intro_comp_quantica}, a construção do algoritmo compreenderia, também, a implementação da operação $\operador{U_f}$, que de certa forma representa a implementação da função $f(x)$. Contudo, isso exigiria um conhecimento prévio desta função! Antes de mais nada, a dúvida que permanece acerca deste paradoxal limbo de recursividade é: se para implementar o algoritmo de Deutsch, com a finalidade de determinar uma característica da função, precisamos de antemão conhecer a função para construir a operação $\operador{U_f}$, qual a necessidade, ou ainda, a utilidade de se implementar o algoritmo?
 
 Este questionamento, apesar de absolutamente pertinente\footnote{Tal questionamento nos faz colocar à prova a finalidade prática do algoritmo de Deutsch. De fato, é possível inferir que este não possui um fim em si mesmo, mas sim, o propósito de servir como parte de algum processo, ou mesmo, como prova de princípios, ou seja, mostrar a eficiência de um computador quântico numa circunstância específica.}, é, para nosso alívio, resultado de um equívoco com relação ao entendimento dos algoritmos de consulta, sua definição e utilidade no contexto da computação quântica. Do que já foi dito até aqui é evidente que o operador Oráculo, que implementa a operação $\operador{U_f}$, de fato tem conhecimento acerca da função $f(x)$, sendo que, para cada função teríamos um mecanismo interno diferente. É evidente, também, que se quiséssemos construir tal operador, precisaríamos, necessariamente, conhecer a função para qual desejamos implementar $\operador{U_f}$. 
 
 Entretanto, a intenção de um algoritmo de consulta é apenas analisar a capacidade de um computador quântico, posto que o operador Oráculo, construído por um terceiro, é fornecido. A complexidade computacional, ou seja, o número de operações lógicas ou tempo de processamento, para um algoritmo de consulta é então definida em termos do número de consultas ao Oráculo.
 
A motivação para a necessidade do operador Oracular fica mais palpável quando nos atentamos para o fato de que, para um processo análogo num computador clássico e com a mesma finalidade do algoritmo de Deutsch, um mecanismo que implementasse $f(x)$ também seria indispensável. A diferença -- além das proprieades da mecânica quântica, é claro -- seria de que, neste caso, para uma função binária com $2$ entradas, seriam necessárias $2$ consultas à esse operador Oráculo. À primeira vista, no entanto, tal motivação não é tão direta pois no contexto clássico esse processo já se tornou um tanto quanto trivial para nós. Uma certa função $f: \mathbb{R} \rightarrow \mathbb{R}$, por exemplo, quando escrita da forma $$f(x)=x^2+3x+1 \, ,$$ define, implicitamente, um algoritmo, ou seja, um conjunto de passos que precisamos seguir a fim de calcular a função para uma determinada entrada $x$, e nesse sentido a ``operação $\operador{U_f}$'' seria implementada da mesma maneira -- seja ela em uma calculadora clássica, ou até mesmo em nossas mentes. Tomando $x=2$, para o exemplo acima, teríamos
 \begin{center}
 	\begin{minipage}{0.4 \textwidth}
 		\textbf{Entrada:} $x=2$\\
 		\textbf{Procedimento:}\\[-0.7cm]
 		\begin{enumerate}
 			\item Multiplicar $2$ por ele mesmo e armazenar o resultado;
 			\item Multiplicar $2$ pelo número $3$, somar com o resultado anterior e armazenar o resultado;
 			\item Adicionar $1$ ao resultado anterior e armazenar;
 			\item Retornar o valor armazenado.
 		\end{enumerate}
 		\vspace{-0.2cm}
 		\textbf{Saída:} $f(x=2) =11$
 	\end{minipage}
 \end{center}

Nesse sentido, supondo que sejamos nós o terceiro agente e que tenhamos a intenção de propor uma disputa entre um computador quântico e um computador clássico, com objetivo de determinar qual é mais rápido em avaliar se uma função binária, com $2$ variáveis de entrada, é constatante ou balanceada. Nós, como agentes externos, iremos determinar qual é a função, construir o operador oráculo, e equipar ambos computadores antes de dar a largada. O computador clássico necessitará $2$ consultas ao oráculo, enquanto o computador quântico, utilizando o algoritmo de Deutsch, necessitará apenas uma consulta, vencendo a disputa. Isso pois, em suma, a essência deste algoritmo é a utilização do paralelismo quântico, em que a operação $\operador{U_f}$ é aplicada uma única vez, e simultaneamente, para todos os possíveis valores de $x$, em superposição. 

Assim, a vantagem de se definir a complexidade do algoritmo em termos de um Oráculo é justamente demonstrar a superioridade do computador quântico fren\-te aos computadores clássicos, na realização de uma cer\-ta tarefa, estando estes nas mesmas condições, munidos do mesmo recurso (o Oráculo), nos privando dos detalhes de parte do processo (a construção e mecanismo interno da operação oracular).

\section{ Algoritmo de Deutsch-Jozsa}
O algoritmo de Deutsh-Josza, resultado de uma junção dos trabalhos de Deutsch e Josza~\cite{DeutschJozsa1992}, se propõe, assim como seu precursor, o algoritmo de Deutsch, a utilizar a propriedade quântica da superposição coerente de estados para obter vantagem em relação à execução de uma mesma tarefa em um dispositivo clássico. Nesse sentido o algoritmo de Deutsch-Josza pode ser entendido como uma generalização do algoritmo de Deutsch, em que, agora, a função $f(x)$ a ser avaliada permite $N$ entradas em seu domínio. Seguindo essa motivação, é possível extrair, então, uma propriedade global desta função $f(x)$, aplicada em $N$ valores, em uma única avaliação, usando o método da computação por paralelismo quântico. Aqui, são necessários $n$ qbits para representar as $N=2^n$ possíveis entradas. Na representação decimal das variáveis, $x$ pode assumir os valores correspondentes aos números naturais de $0$ à $N-1$, sendo sua correspondente representação binária $x_{n-1} \, x_{n-2} \dots x_1 \, x_0$, em que $x_i$ é variável binária que assume valores $0$ ou $1$. Isso implica que
\begin{align}
    x = x_0 \cdot 2^{0} + x_1 \cdot 2^{1}+ x_2 \cdot 2^{2} + \dots + x_{n-1} \cdot 2^{n-1} 
    \, .
\end{align}

De maneira análoga à introduzida na seção \ref{sec:alg_deutsch}, para a construção do algoritmo serão necessários dois registros quânticos, o primeiro com $n = \log_2{N}$ qbits para representar as entradas $x$, e o segundo registro com $1$ qbit para representar os valores possíveis da função $f(x)$, da forma $\ket*{x} \ket*{f(x)}$ com a variável de entrada na representação decimal, ou 
$$\ket*{x_0,x_1,x_2 \dots, x_{n-1}}\ket*{f(x)} \, ,$$
com a variável de entrada na representação binária. Aqui, é importante salientar que as duas representações são equivalentes, e ao longo da construção algébrica do algoritmo, por vezes, será conveniente explicitar uma ou outra. Da mesma forma, é definida, também, a operação unitária $\operador{U_f}$ cuja atuação é dada pela Eq. \eqref{eq:operacao_Uf}.

O sistema é então iniciado com todos os qbits do primeiro registro no estado $\ket*{0}$ e o qbit do segundo registro no estado $\ket*{1}$
\begin{align}
	\label{eq:deutsch-jozsa inicializaçao}
    \ket*{\psi^{(0)}} &= \overbrace{\ket*{0,0,\dots,0}}^{n \text{ vezes}}\ket*{1}
    = \overbrace{ \ket*{0} \otimes \ket*{0} \otimes  \dots  \otimes \ket*{0}}^{n \text{ vezes}} \otimes \ket*{1}
\nonumber \\
&= \underbrace{\ket*{0}^{\otimes n}}_{x} \otimes \underbrace{\ket*{1}}_{f(x)}
    \, .
\end{align}

Novamente, para gerar uma superposição entre todos os estados da base, implementamos portas Hadamard em todos os qbits de ambos os registros (como descrito na seção \ref{sec:intro_comp_quantica}). Assim,
\begin{align}
\label{eq:estado superposicao deutsch}
\ket*{\psi^{(1)}} 
=& \left( \operador{H}^{\otimes n}  \otimes  \operador{H} \right) \ket*{0}^{\otimes n} \ket*{1}
=\left(\operador{H} \ket*{0}\right)^{\otimes n} \otimes \left( \operador{H}\ket*{1} \right)
\nonumber \\
=& \ket*{+}^{\otimes n}  \ket*{-}
\, ,
\end{align}
em que o primeiro registro está em uma superposição, de igual amplitude de probabilidade, de todos os possíveis estados de $x$, da forma
\begin{align}
		\ket*{+}^{\otimes n}
		=& \overbrace{\ket*{+} \otimes \ket*{+} \otimes \dots \ket*{+}}^{n \text{ vezes}}
		\nonumber \\
		=& \frac{\ket*{0}+\ket*{1}}{\sqrt{2}} \otimes\frac{\ket*{0}+\ket*{1}}{\sqrt{2}} \otimes \dots \otimes\frac{\ket*{0}+\ket*{1}}{\sqrt{2}}
		\nonumber \\
		=& \frac{\ket*{00 \dots0} + \ket*{10 \dots0 } + \ket*{01 \dots0} + \dots  + \ket*{11 \dots1} }{\sqrt{2^n}} 
		\nonumber \\ 
		=& \frac{1}{\sqrt{2^n}} \sum_{x_0 = 0}^1 \sum_{x_1 = 0}^1 \dots \sum_{x_{n-1} = 0}^1 \ket*{x_0x_1 \dots x_{n-1}} . \label{eq:estado superposicao deutsch2}
	\end{align} 
	

Aqui, para alívio de notação, será conveniente representar a soma
\begin{align}
	\label{eq:definicao somatoria}
\sum_{x \in \{0,1\}^n} \equiv \sum_{x_0 = 0}^1 ~ \sum_{x_1 = 0}^1 ~ \sum_{x_2 = 0}^1  ~ \dots  ~ \sum_{x_{n-1} = 0}^1
\, ,
\end{align}
e, assim, a Eq. \eqref{eq:estado superposicao deutsch} fica
\begin{align}
	\label{eq:psi1 deutsch-jozsa}
	\ket*{\psi^{(1)}} =  \sum_{x \in \{0,1\}^n}\frac{\ket*{x} }{\sqrt{2^n}} \otimes \frac{\ket*{0}-\ket*{1}}{\sqrt{2}}
	\, .
\end{align}

Dando prosseguimento, aplicando a operação $\operador{U_f}$, ago\-ra para uma função com $N$ entradas, temos
\begin{align}
\ket*{\psi^{(2)}}
=& \operador{U_f} \ket*{\psi^{(1)}}
\nonumber \\
=&\sum_{x \in \{0,1\}^n} \frac{\ket*{x}}{\sqrt{2^{n}}}  \left( \frac{\ket*{0 \oplus f(x)} - \ket*{1 \oplus f(x)}}{\sqrt{2}} \right)
\, .
\end{align}
Considerando que
\begin{align}
\ket*{0 \oplus f(x)} - \ket*{1 \oplus f(x)}
=&
\left\{
\begin{matrix}
\ket*{0} - \ket*{1} & \text{, se } f(x)=0 \\
(-1)\left(\ket*{0} - \ket*{1}\right)  & \text{, se } f(x)=1
\end{matrix}
\right.
,
\end{align}
podemos chegar ao estado final da operação $\operador{U_f}$
\begin{align}
\ket*{\psi^{(2)}} =& \left(\sum_{x \in \{0,1\}^n} (-1)^{f(x)} \frac{\ket*{x}}{\sqrt{2^{n}}}\right) \otimes  \ket*{-}
\, .
\end{align}

Deve-se, agora, aplicar a operação Hadamard no primeiro registro, resultando em
\begin{align}
	\label{eq:psi3 deutsch-jozsa}
	\ket*{\psi^{(3)}}
	=& \left(\operador{H}^{\otimes n}\otimes\operador{I}\right) \ket*{\psi^{(2)}}
	\nonumber \\
	=&\left( \sum_{x \in \{0,1\}^n} (-1)^{f(x)} \frac{\operador{H}^{\otimes n}\ket*{x}}{\sqrt{2^{n}}}\right) \otimes \ket*{-}
	\, .
\end{align}
A partir daqui, omitindo-se o segundo registro
, e desenvolvendo os termos, temos que o termo entre parênteses fica

\begin{widetext}
	\vspace{-0.1cm}
\begin{align}
\label{eq:somatoria deutsch-jozsa}
\sum_{x \in \{0,1\}^n} (-1)^{f(x)} \frac{\operador{H}^{\otimes n}\ket*{x}}{\sqrt{2^{n}}}
=& \frac{1}{\sqrt{2^n}} \sum_{x_0 = 0}^1 \sum_{x_1 = 0}^1  \dots  \sum_{x_{n -1}= 0}^1 (-1)^{f(x_{n-1}\, \dots x_1 \, x_0)} \left(\operador{H} \ket*{x_0} \right) \otimes \left(\operador{H} \ket*{x_1} \right) \otimes \dots \otimes \left(\operador{H} \ket*{x_{n-1}} \right)
\nonumber \\
=& \frac{1}{\sqrt{2^n}} \sum_{x_0 = 0}^1 \sum_{x_1 = 0}^1  \dots  \sum_{x_{n -1}= 0}^1 (-1)^{f(x_{n-1}\, \dots x_1 \, x_0)} \bigotimes_{i=0}^{n-1} \operador{H} \ket*{x_i}
\nonumber \\
=& \frac{1}{\sqrt{2^n}} \sum_{x_0 = 0}^1 \sum_{x_1 = 0}^1  \dots  \sum_{x_{n -1}= 0}^1 (-1)^{f(x_{n-1}\, \dots x_1 \, x_0)} \bigotimes_{i=0}^{n-1} \frac{\ket*{0} + (-1)^{x_i}\ket*{1}}{\sqrt{2}}
\, .
\end{align}

Ainda, podemos reescrever a produtória da Eq. \eqref{eq:somatoria deutsch-jozsa} acima como
\begin{align}
	\label{eq:produtoria deutsch-jozsa}
	 \bigotimes_{i=0}^{n-1} \frac{\ket*{0} + (-1)^{x_i}\ket*{1}}{\sqrt{2}}
	 =&   \frac{\ket*{0} + (-1)^{x_0}\ket*{1}}{\sqrt{2}} \otimes  \frac{\ket*{0} + (-1)^{x_1}\ket*{1}}{\sqrt{2}} \otimes \dots \otimes \frac{\ket*{0} + (-1)^{x_{n-1}}\ket*{1}}{\sqrt{2}}
	 \nonumber \\
	 =& \frac{1}{\sqrt{2^n}} \left(\ket*{0,0,0,\dots,0} + (-1)^{x_0}\ket*{1,0,0,\dots,0} + (-1)^{x_1}\ket*{0,1,0,\dots,0} + (-1)^{x_0+x_1}\ket*{1,1,0,\dots,0} + \dots\right.
	 \nonumber \\
	  &\left.~+ (-1)^{x_0+x_1+\dots x_{n-1}}\ket*{1,1,1\dots,1}\right)
	  \nonumber \\
	 =& \frac{1}{\sqrt{2^n}} \sum_{z_0 = 0}^1 \sum_{z_1 = 0}^1  \dots  \sum_{z_{n-1} = 0}^1 (-1)^{x_0 \cdot z_0 + x_1 \cdot z_1 +
	 	\dots  + x_{n-1} \cdot z_{n-1}} \ket*{z_0,z_1,
	 	\dots ,z_{n-1}}
 	 \nonumber \\
 	=& \frac{1}{\sqrt{2^n}} \sum_{z_0 = 0}^1 \sum_{z_1 = 0}^1  \dots  \sum_{z_{n-1} = 0}^1 (-1)^{\sum_{i=0}^{n-1} x_i\cdot z_i} \ket*{z_0,z_1,
 		\dots ,z_{n-1}}
 	\, ,
\end{align}
com $z_i=0,1$. Assim, substituindo as Eqs. \eqref{eq:somatoria deutsch-jozsa} e \eqref{eq:produtoria deutsch-jozsa} na Eq. \eqref{eq:psi3 deutsch-jozsa}, o vetor de estado $\ket*{\psi^{(3)}}$ fica
\begin{align} 
	\label{eq:estado psi3 final deutsch-jozsa}
\ket*{\psi^{(3)}} =& 
\frac{1}{\sqrt{2^n}} \sum_{x_0 = 0}^1 \sum_{x_1 = 0}^1  \dots  \sum_{x_{n-1} = 0}^1 (-1)^{f(x_{n-1}\, \dots \, x_1 \, x_0)} \frac{1}{\sqrt{2^n}} \sum_{z_0 = 0}^1 \sum_{z_1 = 0}^1  \dots  \sum_{z_{n-1} = 0}^1 (-1)^{\sum_{i=0}^{n-1} x_i\cdot z_i} \ket*{z_0,z_1, \dots ,z_{n-1}} 
\nonumber \\
=& 
\frac{1}{2^n} \sum_{x_0 = 0}^1 \sum_{x_1 = 0}^1  \dots  \sum_{x_{n-1} = 0}^1 ~ \sum_{z_0 = 0}^1 \sum_{z_1 = 0}^1  \dots \sum_{z_{n-1} = 0}^1 (-1)^{\sum_{i=1}^n x_i \cdot z_i + f(x_{n-1}\, \dots \, x_1 \, x_0)} \ket*{z_1,z_2, \dots ,z_n} 
\nonumber \\
=& \sum_{x \in \{0,1\}^n}  \sum_{z \in \{0,1\}^n} \frac{(-1)^{x \cdot z + f(x)} \ket*{z}}{2^n} 
\, ,
\end{align}
em que as somatórias foram escrita conforme a definição na Eq. \eqref{eq:definicao somatoria}, e $x\cdot z \equiv \sum_{i=1}^n x_i \cdot z_i$.
\end{widetext}

Fazendo, ainda, a substituição $c_z \equiv \sum_{x} \frac{(-1)^{x \cdot z + f(x)} }{2^n}$ na Eq. \eqref{eq:estado psi3 final deutsch-jozsa}, o estado $\ket*{\psi^{(3)}}$ fica
\begin{align}
	\label{eq:psi3 deutsch-jozsa cz}
\ket*{\psi^{(3)}}
=& \sum_{z \in \{0,1\}^n}  c_z \ket*{z} \equiv \sum_{z =0}^{N-1} c_z \ket*{z}
\nonumber \\
=& c_0 \ket*{0} + c_1 \ket*{1} + c_2 \ket*{2} +  \dots c_{N-1} \ket*{N-1}
\, .
\end{align}

Agora, analisando a probabilidade de que em uma medição no primeiro registro obtenhamos o estado $\ket*{0}\equiv\ket*{0,0,\dots,0}$, temos
\begin{align}
	\label{eq:coef c0}
	\left|c_0\right|^2 = \left|\frac{1}{2^n}\sum_{x} (-1)^{f(x)} \right|^2
	\, ,
\end{align}
começando a ficar evidente a propriedade de interferência dos estados quânticos. Se $f(x)$ é \textit{constante}, podemos escrever $f(x) = \alpha$ para qualquer $x$, em que $\alpha$ é uma constante arbitrária. Assim, a probabilidade de obtermos o estado $\ket*{0}$ numa medida fica
\begin{align}
\left|c_0\right|^2
=&\left|\frac{(-1)^{\alpha} }{2^n} \sum_{x=0}^{2^n -1} 1 \right|^2
=\left|\frac{(-1)^{\alpha} }{2^n} 2^n \right|^2 =1
\, .
\end{align}
Mas como todas as operações realizadas foram unitárias, da condição de normalização do estado final $\ket*{\psi^{(3)}}$ na Eq. \eqref{eq:psi3 deutsch-jozsa cz} temos
\begin{align}
\sum_{z} \left| c_z \right|^2 &= 1
\nonumber \\
\left| c_0 \right|^2 + \left| c_1 \right|^2 +  \dots + \left| c_{N-1} \right|^2 &=1
\nonumber \\ 
1 + \left| c_1 \right|^2 + \left| c_2 \right|^2 +  \dots  &= 1
\, ,
\end{align}
que só será possível se $c_z = 0$ para qualquer $z \neq 0$. Logo, medindo-se o primeiro registro, necessariamente, obteremos o estado $\ket*{0}$. 

Ainda, se $f(x)$ é \textit{balanceada}, então isso significa que $f(x) = 1$ para metade das entrada $x \in \{0,N-1\}$, e $0$ para a outra metade, de modo que a somatória na Eq. \eqref{eq:coef c0} se anula, e, consequentemente, a probabilidade de obtermos o estado $\ket*{0}$ numa medida fica
\begin{align}
	\left|c_0\right|^2 =0
	\, ,
\end{align}
implicando que $c_0=0$. Logo, medindo-se o primeiro registro, pelo menos um dos $n$ qbits estará, necessariamente, no estado $\ket*{1}$, e obteremos qualquer outro estado que não $\ket*{0}$.

\section{Implementação do algoritmo de Deutsch-Jozsa}

Em essência, o algoritmo de Deutsch-Jozsa se beneficia do paralelismo quântico na aplicação da operação $\operador{U_f}$, gerando um estado em que cada vetor de estado que representa uma entrada $x$ está emaranhado com o vetor de estado que representa o valor da função $f(x)$ para aquela entrada, da forma 
\begin{align}
	\frac{1}{\sqrt{N}} \sum_{x=0}^{N-1} \ket*{x} \xrightarrow{~~ \operador{U_f}~~} \frac{1}{\sqrt{N}} \sum_{x=0}^{N-1} \ket*{x}\ket*{f(x)}
	\, .
\end{align}
Ainda, é feito com que esses estados se interfiram destrutivamente, levando a um estado final cuja medida fornece o resposta para o problema. 

O objetivo desta seção é mostrar, passo a passo, a implementação do algoritmo de Deutsch-Jozsa utilizando o computador quântico da IBM. Aqui, a intenção é utilizar o algoritmo como prova de princípio, e não como fim em si mesmo, ou seja, ele será utilizado não para avaliar de maneira eficiente uma função desconhecida, mas sim para demonstrar a eficiência do computador quântico, frente aos computadores clássicos, na tarefa de se avaliar uma determinada função. Nesse sentido, nós faremos o papel do terceiro agente, como mencionado na seção \ref{sec: caixa preta}, selecionando a função, preparando a respectiva operação oracular $\operador{U_f}$, e equipando o computador quântico.

Serão preparados exemplos para duas funções, uma constante e outra balanceada, sendo ambas da forma $f: \{0,1\}^3 \rightarrow \{0,1\}$, ou seja, $N=2^n=8$ possíveis entradas, com $n=3$ qbits para codificar essas entradas. Assim, serão necessários: (i) Um registro quântico com 3 qbits para codificar as variáveis de entrada $x$; (ii) um registro quântico com 1 qbit para armazenar $f(x)$; e (iii) Um registro clássico com 3 bits para a medida final do registro das entradas. Ainda, para cada função, iremos utilizar determinadas portas lógicas para simular a respectiva operação oracular. O código \textit{QASM} que implementa o algoritmo, elucidado a seguir, está disponível em \cite{qasm_codes}.

\subsection{Deutsch-Jozsa para uma função constante}

\begin{figure}
	\centering
	\includegraphics[width = 0.6 \columnwidth]{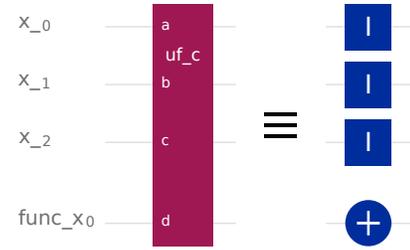}
	\caption{Simulação do mecanismo interno da operação $\operador{U_f}$ para a função constante $f_c(x)$ selecionada. A operação, definida a partir das portas lógicas quânticas mostradas, implementa a função $f_c(x)$.}
	\label{fig:operador_Uf_func_cte}
	\vspace{-0.4cm}
    \begin{flushleft}
	{\small Fonte: IBM Q~\cite{IBMQ} e editada pelos autores.}
    \end{flushleft}
\end{figure}

Consideremos, inicialmente, uma função constante $f_c: \{0,1\}^3 \rightarrow \{0,1\}$ tal que $f_c(x)=1$ para qualquer $x \in \{0,7\}$, cuja tabela \ref{tab:tabela verdade func cte} relaciona os valores de saída da função para as possíveis entradas, e a figura \ref{fig:operador_Uf_func_cte} mostra uma simulação da operação oracular, que implementa a função $f_c(x)$, utilizando-se portas lógicas quânticas.

\begin{table}[ht]
	\centering
	\begin{tabular}{|c|ccc|c|}
		\hline
		\multicolumn{4}{|c|}{\textbf{Entrada}} & \textbf{Saída} \\ \hline
		$x$    & $x_2$    & $x_1$    & $x_0$   & $f_c(x)$         \\ \hline
		0      & 0        & 0        & 0       & 1              \\
		1      & 0        & 0        & 1       & 1              \\
		2      & 0        & 1        & 0       & 1              \\
		3      & 0        & 1        & 1       & 1              \\
		4      & 1        & 0        & 0       & 1              \\
		5      & 1        & 0        & 1       & 1              \\
		6      & 1        & 1        & 0       & 1              \\
		7      & 1        & 1        & 1       & 1              \\ \hline
	\end{tabular}
\caption{Tabela verdade da função constante $f_c(x)$, com as possíveis entradas (em representação decimal no primeiro bloco, e em representação binária, no segundo bloco) com os respectivos valores da função.}
\label{tab:tabela verdade func cte}
\end{table}

\begin{figure}
    \centering
    \includegraphics[width = \columnwidth]{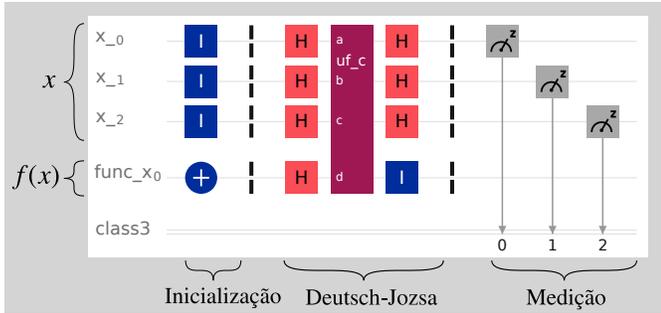}
    \caption{Sequência de operações lógicas necessária para a implementação do algoritmo de Deustch-Jozsa. Na inicialização, os registros quânticos, configurados todos no estado$\ket*{0}$ (padrão), são colocados no estado indicado na Eq. \eqref{eq:deutsch-jozsa inicializaçao}. Aqui foi utilizada a operação oracular $\operador{U_f}$ simulada para implementar a função constante $f_c(x)$.}
    \label{fig:implementaçao funcao cte}
    \vspace{-0.4cm}
\begin{flushleft}
	{\small Fonte: IBM Q~\cite{IBMQ}  e editada pelos autores.}
\end{flushleft}
\end{figure}

\begin{figure}
    \centering
    \includegraphics[width= \columnwidth]{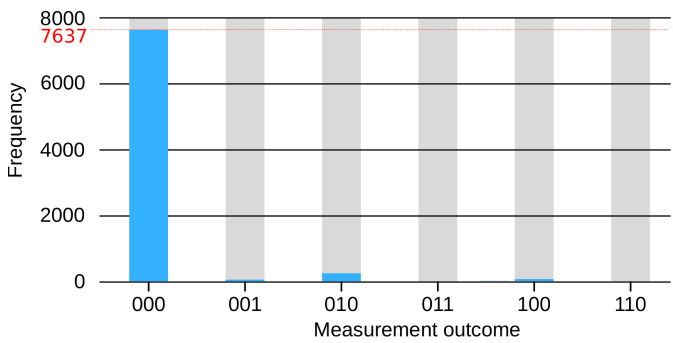}
    \caption{Histograma dos resultados da realização de 8000 experimentos, identicamente preparados no computador quântico IBMQ Manila, na execução do algoritmo de Deutsch-Jozsa para a função constante $f_c(x)$ simulada.}
    \label{fig:histograma funçao cte}
    \vspace{-0.4cm}
\begin{flushleft}
	{\small Fonte: IBM Q~\cite{IBMQ} e editada pelos autores.}
\end{flushleft}
\end{figure}

Assim, equipando o computador quântico com a operação oracular, e implementando o algoritmo, conforme mostrado na figura \ref{fig:implementaçao funcao cte}, obtemos os resultados, esquematizados no histograma da figura \ref{fig:histograma funçao cte}. Da análise deste, o resultado mais notório é o de que em $\approx 95\%$ dos experimentos encontramos como saída do algoritmo o estado $\ket*{x_2 \, x_1 \, x_0} = \ket*{000}$. Ainda, observamos a ocorrência de probabilidades não nulas para resultados não esperados, relacionadas aos erros intrínsecos ao computador quântico, conforme mencionado ao final de Sec. \ref{sec:ibmq_experience}. A partir do exame destes resultados, assumindo certa condescendência com relação aos erros citados\footnote{Apesar de inevitáveis, os erros intrínsecos aos computadores quânticos podem ser minimizados. Protocolos de mitigação de erros podem ser encontrados na documentação fornecida pela plataforma \cite{IBMQ_documentation,Qiskit}.} e nos atendo à interpretação do resultado mais expressivo ($\approx 95\%$ de ocorrência do estado $\ket{000}$), podemos confirmar a ação do algoritmo de Deutsch na determinação de uma função constante, em apenas uma única medida.

\subsection{Deutsch-Jozsa para uma função balanceada}

\begin{figure}
	\centering
	\includegraphics[width=0.6 \columnwidth]{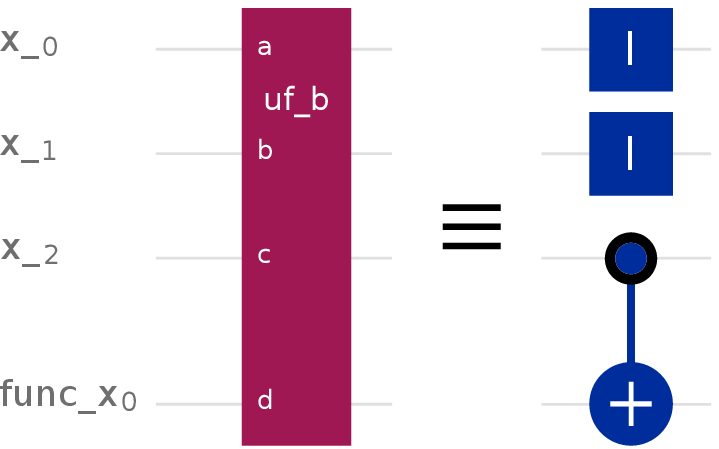}
	\caption{Simulação do mecanismo interno da operação $\operador{U_f}$ para a função balanceada $f_b(x)$ selecionada. A operação, definida a partir das portas lógicas quânticas mostradas, implementa a função $f_b(x)$.}
	\label{fig:operador_Uf_func_balanc}
	\vspace{-0.4cm}
\begin{flushleft}
	{\small Fonte: IBM Q~\cite{IBMQ} e editada pelos autores.}
\end{flushleft}
\end{figure}

Consideremos, agora, uma função balanceada $f_b:\{0,1\}^3$ $\rightarrow \{0,1\}$ tal que 
\begin{align}
	\label{eq:funçao balanceada}
	f_b(x) = 
	\left\{
	\begin{aligned}
	0 & \text{ se $x \leq 3$,}\\
	1 & \text{ se  $x > 3$,}
	\end{aligned}
	\right.
\end{align}
cuja tabela \ref{tab:tabela verdade func balanc} relaciona os valores de saída da função para as possíveis entradas, e a figura \ref{fig:operador_Uf_func_balanc} mostra uma simulação da operação oracular, que implementa a função $f_b(x)$, utilizando-se portas lógicas quânticas.

\begin{table}[ht]
	\centering
	
	\begin{tabular}{|c|ccc|c|}
		\hline
		\multicolumn{4}{|c|}{\textbf{Entrada}} & \textbf{Saída} \\ \hline
		$x$    & $x_2$    & $x_1$    & $x_0$   & $f_b(x)$         \\ \hline
		0      & 0        & 0        & 0       & 0              \\
		1      & 0        & 0        & 1       & 0              \\
		2      & 0        & 1        & 0       & 0              \\
		3      & 0        & 1        & 1       & 0              \\
		4      & 1        & 0        & 0       & 1              \\
		5      & 1        & 0        & 1       & 1              \\
		6      & 1        & 1        & 0       & 1             \\
		7      & 1        & 1        & 1       & 1              \\ \hline
	\end{tabular}
	\caption{Tabela verdade da função balanceada $f_b(x)$, com as possíveis entradas (em representação decimal no primeiro bloco, e em representação binária, no segundo bloco) com os respectivos valores da função, no terceiro bloco.}
	\label{tab:tabela verdade func balanc}
	\vspace{-0.4cm}
\end{table}

\begin{figure}
	\centering
	\includegraphics[width = \columnwidth]{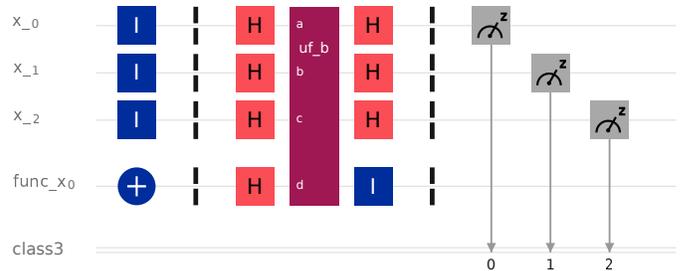}
	\caption{Esquema análogo ao da figura \ref{fig:implementaçao funcao cte},  com as operações lógicas necessárias para a implementação do algoritmo de Deustch-Jozsa, utilizando, aqui, a operação oracular $\operador{U_f}$ simulada para implementar a função balanceada $f_b(x)$.}
	\label{fig:implementaçao funcao balanc}
	\vspace{-0.4cm}
\begin{flushleft}
	{\small Fonte: IBM Q~\cite{IBMQ} e editada pelos autores.}
\end{flushleft}
\end{figure}

\begin{figure}
	\centering
	\includegraphics[width=\columnwidth]{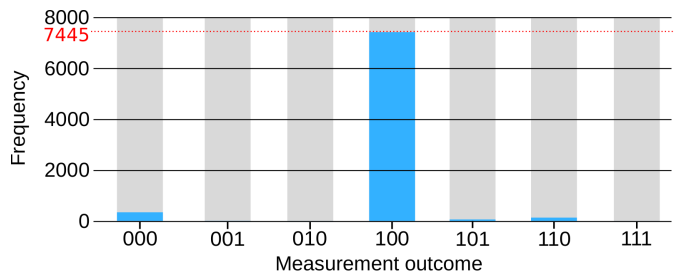}
	\caption{Histograma dos resultados da realização de 8000 experimentos, identicamente preparados no computador quântico IBMQ Manila, na execução do algoritmo de Deutsch-Jozsa para a função balanceada $f_b(x)$ simulada.}
	\label{fig:histograma funçao balanc}
	\vspace{-0.4cm}
\begin{flushleft}
	{\small Fonte: IBM Q~\cite{IBMQ} e editada pelos autores.}
\end{flushleft}
\end{figure}

Assim, equipando o computador quântico com a operação oracular, e implementando o algoritmo, conforme mostrado na figura \ref{fig:implementaçao funcao balanc}, obtemos os resultados, esquematizados no histograma da figura \ref{fig:histograma funçao balanc}.  Novamente, percebemos que o resultado mais expressivo mostra a obtenção do estado final $\ket*{x_2 \, x_1 \, x_0} = \ket*{100}$ em $\approx 93\%$ dos 8000  experimentos, além da ocorrência de probabilidades não nulas para resultados não esperados. Outra vez, sendo condescendentes com os erros intrínsecos, vemos que, como pelo menos um dos 3 qbits se encontra num estado diferente de $\ket*{0}$, podemos atestar a performance do algoritmo de Deutsch na determinação de uma função balanceada, em apenas uma única medida.

\section{Conclusão}

Este artigo teve como propósito introduzir os fundamentos e aspectos básicos da computação quântica, dan\-do como exemplos os algoritmos de Deutsch e Deutcsh-Joz\-sa, tanto do ponto de vista teórico, com suas etapas matemáticas detalhadas nas seções anteriores, quanto do ponto de vista experimental, a partir da implementação nos computadores quânticos da IBM, disponíveis gratuitamente na plataforma IBM Quantum Experience. Também, para uma melhor compreensão dos algoritmos mencionados, foi dado um importante enfoque às características dos \textit{Quantum Query Algorithms} e ao mecanismo de caixa preta. 

Como prova de princípio, o algoritmo de Deutsch-Joz\-sa foi implementado, o que permitiu constatar a superioridade do computador quântico na avaliação das características de uma função. Tal avaliação pôde ser realizada em apenas uma única medição, o que não se faz possível quando utiliza-se um computador clássico. O código em linguagem QASM utilizado para a implementação dos algoritmos estão disponíveis em um repositório online no GitHub \cite{qasm_codes}. Por meio dele e da criação de uma conta gratuita na plataforma da IBM Quantum Experience, é possível verificar experimentalmente as principais características e vantagens da computação quântica.

\section*{Agradecimentos}
Este trabalho teve o apoio do Conselho Nacional de Desenvolvimento Científico e Tecnológico (CNPq), da Coordenação de Aperfeiçoamento de Pessoal de Nível Superior (CAPES) - Código 001, e do Instituto Nacional de Ciência e Tecnologia para Informação Quântica (INCT-IQ/CNPq), Processo No. 465469/2014-0. C.J.V.-B. também agradece o apoio da Fundação de Pesquisa do Estado de São Paulo (FAPESP), Processo No. 2019/11999-5, e do CNPq, Processo No. 307077/2018-7. Queremos agradecer ao Grupo de Óptica Quântica e Informação Quântica do Departamento de Física (DF) da Universidade Federal de São Carlos (UFSCar) por todas as ricas discussões, que contribuíram para esse trabalho. E.V.B.O também agradece à Suzana V.B. de Oliveira por sua assistência e considerações à respeito dos aspectos que trariam maior valor didático ao trabalho, para um público não especializado.


\end{document}